\documentclass[12pt]{article}
\usepackage{latexsym}
\def\dslash {\partial\!\!\!/}
\newcommand{\be}{\begin{equation}}
\newcommand{\ee}{\end{equation}}
\newcommand{\no}{\noindent}
\newcommand{\bear}{\begin{eqnarray}}
\newcommand{\ear}{\end{eqnarray}}
\newcommand{\ba}{\begin{eqnarray*}}
\newcommand{\ea}{\end{eqnarray*}}

\newcommand{\F}{{\cal F}}
\newcommand{\Z}{\mbox{\boldmath${\cal Z}$}}

\newcommand{\g}{{\mathsf{g}}}

\newcommand{\mes}{\mbox{\boldmath$d \mu$}}

\newcommand{\opOmega}{\mbox{\boldmath$\Omega$}}
\newcommand{\optheta}{\mbox{\boldmath$\theta$}}

\newcommand{\opomega}{\mbox{\boldmath$\omega$}}
\newcommand{\copomega}{{\check{\mbox{\boldmath$\omega$}}}}

\newcommand{\G}{{\mathbf{G}}}
\newcommand{\bphi}{\mbox{\boldmath$\phi$}}
\newcommand{\tbphi}{\widetilde{\mbox{\boldmath$\phi$}}}

\title{The Thirring-Wess Model Revisited: A Functional Integral Approach}
\author{L. V. Belvedere* and A. F. Rodrigues**\\
\small
{ \it Instituto de F\'{\i}sica - Universidade Federal Fluminense}\\
\small
{\it Av. Litor\^anea, S/N, Boa Viagem, Niter\'oi,
CEP.24210-340}\\
\small{\it Rio de Janeiro - Brasil}}

\date{\today}

\begin{document}
%\twocolumn
\maketitle
%%%%%%%%%%%%%%%%%%%%%%%%%%%%%%%%%%%%%%%%%%%%%%%%%%%%%%%%%%%%%%%
%\vspace{0.5cm}

\begin{abstract}
We consider the Wess-Zumino-Witten theory to obtain
the functional integral bosonization of the Thirring-Wess model with 
an arbitrary regularization parameter. Proceeding a systematic of 
decomposing the Bose field algebra into gauge-invariant- and 
gauge-noninvariant field subalgebras, we obtain the local decoupled 
quantum action. The generalized operator solutions for the equations 
of motion are reconstructed from the functional integral 
formalism. The isomorphism between the $QED_2$ ($QCD_2$) with broken 
gauge symmetry by a regularization prescription and the Abelian (non-Abelian) 
Thirring-Wess model with a fixed bare mass for the meson field is established.

\footnote{*e-mail: belve@if.uff.br}
\footnote{**e-mail: armflavio@if.uff.br}
\footnote{\bf${\cal L}$atex file:aftw.tex}
\end{abstract}
\vspace{0.5cm}
%%%%%%%%%%%%%%%%%%%%%%%%%%%%%%%%%%%%%%%%%%%%%%%%%%%%%%%%%%%%%%%%%%%%%%

%%%%%%%%%%%%%%%%%%%%%%%%%%%%%%%%%%%

\section{Introduction}

The Thirring-Wess (TW) model \cite{TWA} was considered in 
Refs. \cite{TWB,TWC,TWD,TWE,LS,RS} in order to investigate the way in 
which $QED_2$ can be understood as a limit of a vector meson theory 
when the bare mass ($m_o$) of the meson tends to zero. The standard 
TW model, in which the gauge invariant regularization prescription is 
adopted, corresponds to quantize the model absorbing all effects of 
the gauge symmetry breakdown in the bare mass of the meson theory. The 
computation of the vector current is performed using the gauge invariant 
regularization for the point-splitting limiting procedure, which 
corresponds to the regularization parameter $a = 2$. Within a 
formulation in a positive-definite metric Hilbert 
space, the limit $m_o \rightarrow 0$ does not exist for 
the Fermi field operator ($\psi$) and vector field operator  (${\cal A}_\mu$) 
themselves \cite{RS}. The zero mass limit is not well defined for the general 
Wightman functions that provide a representation of the gauge noninvariant 
field algebra of the TW model. However, the gauge invariant Wightman functions 
of the $QED_2$ are obtained as the zero mass limit of the Wightman functions 
of the gauge invariant field subalgebra of the standard TW model. 

More recently, the $QED_2$ with broken gauge symmetry, by the use of
an arbitrary regularization parameter $a$, has been considered in 
Refs.\cite{Tiao} and entitled ``anomalous'' \textbf{vector} Schwinger model. The 
standard $QED_2$ should corresponds to the gauge invariant 
regularization $a = 2$. However, also in this case, the
gauge invariant limit $a \rightarrow 2$ does not exist for the 
fields $\psi$ and ${\cal A}_\mu$ themselves. The limit $a \rightarrow 2$ is 
well defined only for the gauge invariant subset of Wightman 
functions. The structural problem associated with the zero bare 
mass limit $m_o \rightarrow 0$ of the Wightman functions for the standard 
TW model can be mapped into the corresponding problem of perform the 
limit $a \rightarrow 2$ in the $QED_2$ with gauge symmetry breakdown. 

One of the purposes of the present work is to fill a gap in the 
existing literature by discussing structural aspects of the TW model 
from the functional integral formulation and by extending the 
analysis to the general TW model regularized with an arbitrary 
parameter $a$. To this end, we 
use the Abelian reduction of the Wess-Zumino-Witten theory to consider 
the functional integral bosonization of the generalized TW model. In order 
to obtain a better insight into the behaviour of gauge variant 
operators, in the present approach we addopt the systematic of
decomposing step-by-step the Bose field algebra into gauge-invariant- (GI) 
and gauge-noninvariant (GNI) field subalgebras, such that the 
effective bosonized quantum action decouples into GI and GNI pieces. The 
gauge symmetry breakdown is characterized by the presence of a 
non-canonical free massless Bose field. The vector field with 
bare mass $m_o$ acquires a dynamical mass

\be
\tilde m_o^2 = \frac{e^2}{4 \pi}\,( a + 2 )\,+\,m_o^2\,.
\ee

\no Within the present approach we obtain a formulation in an 
indefinite-metric Hilbert space of states and the Proca equation is
satisfied in the weak form. The generalized field operators are 
reconstructed from the functional integral formalism and are written 
as gauge transformed fields by an operator-valued gauge 
transformation. In the indefinite-metric formulation, the GI limit 
can be performed and we obtain the corresponding field operators of $QED_2$, as
obtained by Lowenstein-Swieca \cite{LS}. Performing a canonical 
transformation, the singular gauge part becomes the identity and we 
obtain the generalized operator solution for the coupled Dirac-Proca 
equations. For the gauge invariant regularization $a = 2$, we recover 
the operator solution of the standard TW model, as obtained by 
Lowenstein-Rothe-Swieca \cite{LS,RS}. Since 
in this case the longitudinal current does not carry any fermionic 
charge selection rule, commutes with itself and commutes with all 
operators belonging to the field algebra, it is reduced to 
the identity operator. This leads to a positive-metric Hilbert space of states 
and the coupled Dirac-Proca equations are satisfied in the strong form. Since
in this case the Fermi field operator is given in terms of the 
charge-carrying fermion operator of the Thirring model, the zero mass limit
exists only for the GI field subalgebra.  This streamlines the 
presentation of Refs. \cite{LS,RS}. 

Another purpose of the present paper is to discuss the isomorphism between 
the $QED_2$ with broken local gauge invariance ( the ``anomalous'' \textbf{vector} 
Schwinger model considered in \cite{Tiao}) and the TW model. Using the 
Wess-Zumino-Witten theory, we show that the effective quantum action of the
$QED_2$ quantized with a gauge noninvariant regularization $b \neq 2$, is
equivalent to the quantum action of the TW model with a 
vector field with bare mass

\be
m_o^2 = \frac{e^2}{4 \pi}\,( b - a )\,,
\ee

\no where $a < b$ is the parameter used in the regularization of the 
TW model. In this way, the gauge invariant limit $b \rightarrow 2$ for 
the $QED_2$ with broken gauge symmetry, is mapped into the 
limit $m_o \rightarrow 0$ for the standard TW model ($a = 2$). As is well 
known \cite{RS,S}, the confinement phenomenon in the standard $QED_2$ is
associated with the absence of charge sectors.  In this way, the
conclusion of Ref. \cite{Tiao}, according with the parameter $a$ controls the
screening and confinement properties is nothing but that the TW model exhibits
a charge-carrying fermion operator and thus there is no confinement.  

The introduction of the mass term for the Fermi field of the generalized 
TW model is also considered. For the GI regularization we recover the operator
solution obtained by Rothe-Swieca \cite{RS} for the massive TW model.

In the last section we consider the functional integral bosonization of the
non-Abelian TW model with an arbitrary regularization parameter. In this 
case, the quantum action of the non-Abelian TW model is mapped into 
the action of the $QCD_2$ with gauge symmetry breakdown.

%%%%%%%%%%%%%%%%%%%%%%%%%%%%%%%%%%%%%%%%

\section{Functional Integral Bosonization}

%%%%%%%%%%%%%%%%%%%%%%%%%%%%%%%%%%%%%%%%

\setcounter{equation}{0}

The Thirring-Wess (TW) model is defined \cite{TWA} 
from the classical Lagrangian density  \footnote{Our conventions are: 
$ g^{00} = 1 $, $\epsilon^{01} = - 
\epsilon_{01} = - \epsilon^{1 0} = 1$, $\gamma^\mu \gamma^5 = 
\epsilon^{\mu \nu} \gamma_\nu$, $ \dslash = \gamma^\mu \partial_\mu;$, 
$\tilde \partial_\mu = \epsilon_{\mu \nu} \partial^\nu$, 
$\gamma^5 = \gamma^0 \gamma^1$; 
%$\psi = \pmatrix{\psi_{_{\!\ell}} 
%\cr \psi_{_{\!r}}}$;
$\gamma^0 = \pmatrix{0 & 1 \cr 1 & 0}, \gamma^1 = 
\pmatrix{0 & 1 \cr - 1 & 0}$; $\partial_\pm = \partial_0 \pm \partial_1$; 
$ {\cal A}_\pm = {\cal A}_0 \pm {\cal A}_1$;\,$\dslash = 
\gamma^\mu \partial_\mu$; For a free massless scalar 
field, $\tilde \partial_\mu \,\tilde \Phi\,=\,-\,\partial_\mu\,
 \Phi$. The Fermi field $\Psi$ is the two-component 
 spinor $\Psi = \pmatrix{\psi_1 \cr \psi_2}$\,.}
, 

\be\label{lagtw}
{\cal L} = -\,\frac{1}{4}\,{\cal F}_{\mu \nu}{\cal F}^{\mu \nu}\,+\,
\bar \psi\,(\,i\,\gamma^\mu\,\partial_\mu\,-\,e\,
\gamma^\mu\,{\cal A}_\mu\,)\,\psi\,+\,\frac{1}{2}\,m_o^2\,{\cal A}_\mu {\cal A}^\mu\,,
\ee

\noindent where the field-strenght tensor  $\F_{\mu \nu}$ is given by

\be
\F_{\mu \nu} = \partial_\nu {\cal A}_\mu - \partial_\mu {\cal A}_\nu\,.
\ee

\noindent The local gauge invariance is broken by the mass term for 
the vector field. In the classical level, the local gauge invariance can be  
restored by performing in (\ref{lagtw}) the limit $m_o \rightarrow 0$. This 
limit corresponds to the classical Lagrangian of the two-dimensional 
electrodynamics. However, in the quantum level, the zero mass limit for 
the vector field, even for a gauge invariant regularization prescription, is 
well defined only for the gauge invariant subset of 
Wightman functions \cite{RS}.

%%%%%%%%%%%%%%%%%%%%%%%%%%%%%%%%%%%%%%%%%%%%%%%
\subsection{Decoupled  Quantum Action}
%%%%%%%%%%%%%%%%%%%%%%%%%%%%%%%%%%%%%%%%%%%%%%%%

In order to obtain the bosonized action, we shall consider the Abelian
reduction of the Wess-Zumino-Witten theory \cite{WZW}. To this end, let 
us consider the generating functional (in Minkowski space),

\be\label{gf}
{\Z}\,[\,\bar \vartheta, \vartheta, \jmath^\mu\,]\,=\,
\Big \langle\,
e^{\,i\,\int\,d^2 z\,(\,\bar \vartheta\,
\psi\,+\,\bar \psi\,\vartheta\,+\,\jmath_\mu\,{\cal A}^\mu\,)} \Big \rangle\,,
\ee

\noindent where the average is taken with respect to the functional integral
measure,

\be\label{mes}
\mes\,=\,
\int\,{\cal D}\,{\cal A}_\mu\,{\cal D}\,\bar \psi\,{\cal D}\,\psi\,
e^{\,i\,S [\bar \psi, \psi, {\cal A}_\mu]}\,,
\ee

\noindent and the Lagrangian density is written as, 

\be\label{lagtw1}
{\cal L}\,=\,-\,\frac{1}{4}\,{\cal F}_{\mu \nu}\,{\cal F}^{\mu \nu}\,+\,
\psi_1^\dagger\,D_+ ( {\cal A}\,)\,\psi_1\,+\,
\psi_2^\dagger\,D_- ({\cal A})\,\psi_2\,+\,
\frac{1}{2}\,m_o^2\,{\cal A}_+ {\cal A}_-\,,
\ee

\noindent where

\be
D_\pm ({\cal A} ) \doteq (\,i\,\partial_\pm\,-\,e\,{\cal A}_\pm\,)\,.
\ee

The vector field can be parametrized in terms of the 
$U (1)$ group-valued Bose fields $(U,V)$ as follows \cite{BRRS,BRR,Bel}

\be\label{cva1}
{\cal A}_+\,=\,-\,\frac{1}{e}\,U^{-1}\,i\,\partial_+\,U\,,
\ee

\be\label{cva2}
{\cal A}_-\,=\,-\,\frac{1}{e}\,V\,i\,\partial_-\,V^{-1}\,,
\ee

\noindent with

\be\label{U}
U \,=\,e^{\textstyle\,2\,i\,\sqrt \pi\,u }\,,
\ee

\be\label{V}
V \,=\,e^{\textstyle\,2\,i\,\sqrt \pi\,v}\,,
\ee

\noindent such that,

\be
\bar \psi\,\slash\!\!\!\!D ({\cal A})\, \psi \,=\,
(U \psi_1 )^\dagger\,(i\,\partial_+)\,(U \psi_1 )\,+
(V^{-1} \psi_2 )^\dagger\,(i\,\partial_-)\,(V^{-1} \psi_2 )\,.
\ee

\no In order to decouple the Fermi- and vector fields in the 
Lagrangian (\ref{lagtw}), the spinor components $(\psi_1,\psi_2)$, are 
parametrized in terms of the Bose fields $(U,V)$ and the free Fermi field 
components $(\chi_1,\chi_2)$,

\be \label{fr1}
\psi_1\,=\,U^{-1}\,\chi_1\,,
\ee

\be \label{fr2}
\psi_2\,=\,V\,\chi_2\,.
\ee

In order to introduce the systematic of decomposing the Bose field 
algebra $\Im^B$,

\be
\Im^B = \Im^B \{U,V\}\,,
\ee

\noindent  into gauge invariant (GI) and gauge noninvariant (GNI) field 
subalgebras, let us consider a {\it class of local gauge transformations} ${g}$ acting 
on the Bose fields $U, V$, as follows,

\be\label{gtu}
{g}\, :\, U\,\rightarrow\,^{g} U \,=\,U\,{g}\,,
\ee

\be\label{gtv}
{g}\,:\,V\,\rightarrow \, ^{g} V\,=\,{g}^{-1}\,V\,,
\ee

\noindent with

\be
{g} (x) = e^{\,\textstyle\,2\,i\,\sqrt \pi\,{\Lambda}\, (x)}\,.
\ee

\noindent Under ${g}$-transformations the fields $(u,v)$ transform according
with,

\be
{g}\,:\,u\,\rightarrow\,u + \Lambda\,,
\ee

\be
{g}\,:\,v\,\rightarrow\,v - \Lambda\,.
\ee

\noindent The field combination $( u + v ) $ is gauge invariant, whereas the
combination $ ( u - v ) $ is gauge noninvariant. The vector field and 
the Fermi field transform under ${g}$ according with,

\be
^{g} {\cal A}_\mu = {\cal A}_\mu + \frac{1}{e}\,{g}\,i\,\partial_\mu\,
{g}^{-1}\,,
\ee

\be
^{g} \psi_1 = \,(\,^{g}U^{-1}\,)\,\chi_1\,=\,{g}^{-1} 
\psi_1\,,
\ee

\be
^{g} \psi_2 = \,(\,^{g}V\,)\,\chi_2\,=\,{g}^{-1}\,
\psi_2\,.
\ee

\noindent Let us introduce the gauge invariant Bose field ${\G}$,

\be \label{G}
{\G} \doteq U V\,=\,e^{\textstyle \,2\,i\,\sqrt \pi\,
( u + v )}\,, 
\ee

\noindent such that,

\be
^{g} {\G}\,=\,^{g}U\,^{g}V\,=\,U V\,=\, {\G}\,.
\ee

\noindent In terms of the field ${\G}$, the field-strenght tensor is given by,

\be\label{F}
{\cal F}_{01} = \frac{1}{2}\,(\partial_- {\cal A}_+ - \partial_+ {\cal A}_- )\,
=\,-\,\frac{1}{2 e}\,\partial_+ ( {\G}^{-1}\,i\,\partial_- {\G} )\,
=\,\frac{\sqrt \pi}{e}\,\Box ( u + v )\,.
\ee

\noindent and the Maxwell action can be written as,

$$
S_M [ UV ] = 
\frac{1}{2\,\tilde \mu_o^2}\,\int\,d ^2 z\,(u\,+\,v)\,\Box^2\,
(u\,+\,v ) =
$$

$$
S_M [ {\G} ] = 
\frac{1}{8 e^2}\,\int\,d ^2 z\,[\partial_+( {\G}\,i\, 
\partial_- {\G} ^{-1})]^2
$$

\be\label{ma}
=\,\frac{1}{8 e^2}\,\int\,d ^2 z\,[\partial_-( {\G}^{-1}\,i\, 
\partial_+ {\G} )]^2\,,
\ee

\noindent where we have defined 

\be
\tilde \mu_o^2\,=\,\frac{e^2}{4\,\pi}\,.
\ee

Let us return to the functional integration in the generating 
functional. Introducing the identities,

\be
1\,=\,\int\,{\cal D} U\,[ det {\cal D}_+ (U)]\,
\delta ( e  {\cal A}_+ - U^{-1}\,i\,\partial_+ U )\,,
\ee

\be
1\,=\,\int\,{\cal D} V\,[ det {\cal D}_- (V)]\,
\delta ( e  {\cal A}_- - V\,i\,\partial_- V^{-1} )\,,
\ee

\noindent the change of variables $\{{\cal A}_+,{\cal A}_-\} \,
\rightarrow\,\{U, V\}$ is performed by integrating over the vector field 
components ${\cal A}_\pm$. Performing the Fermi field 
rotations (\ref{fr1})-(\ref{fr2}), and taking 
due account of the Jacobians in the change of the
integration measure in the generating functional, we obtain 
the effective integration measure \cite{BRRS,BRR,Bel},

\be
\mes\,=\,{\cal D}\,\bar \chi\,{\cal D}\,\chi\,
{\cal D}\,U\,{\cal D}\,V\,
e^{\,i\, S [U,V,\bar \chi,\chi]}\,e^{\,i\, S^\prime [U,V,a,m_o]}\,,
\ee

\noindent where,

\be
S [U,V,\bar \chi,\chi]\,=\,
\int\,d ^2 z\,\bar \chi\,i\,\dslash\,\chi\,
+\,S_M [UV]\,,
\ee

\no and $S^\prime [U,V,a,m_o]$ is the GNI contribution, which is 
given in terms of the Wess-Zumino-Witten
(WZW) functionals $\Gamma [U], \Gamma [V]$,

\be\label{JRa}
S^\prime [U,V,a,m_o]\,=\,-\,\Gamma [U]\,-\,\Gamma [V]\,-\,\frac{1}{2\,e^2}\,( a\,\tilde \mu_o^2\,+\,m_o^2 )\,
\int\,d ^2 z\,\Big (U^{-1}\,\partial_+ U \Big )\,
\Big ( V\,\partial_- V^{-1} \Big )\,.
\ee

\no The term carrying the regularization parameter $a$ in the action 
(\ref{JRa}) corresponds to the Jackiw-Rajaraman action,

\be
S_{JR} \,=\,\frac{a}{2}\,\tilde \mu_o^2\,\int\,d\,^2 z\,{\cal A}_+\,{\cal A}_-\,,
\ee

\no  which in an anomalous model, such as chiral $QED_2$, characterizes the 
quantization ambiguity \cite{JR}. The value $a = 2$ corresponds to the 
GI regularization.  The WZW functionals 
enter (\ref{JRa}) with negative level \cite{WZW}. In the Abelian case, the WZW 
functional reduces to the action of a free massless scalar field,

\be
\Gamma [ h ]\,=\,\frac{1}{8 \pi}\,\int\,d ^2 z\,( \partial_\mu\,h)\,
( \partial^\mu\,h^{-1} )\,.
\ee

\no Using the Abelian reduction of the 
Polyakov-Wiegmann (PW) identity \cite{PW},

\be
\Gamma [ g h ]\,=\,\Gamma [ g ] + \Gamma [ h ]\,+\,\frac{1}{4 \pi}\,
\int\,d ^2 z\,( g^{-1} \partial_+ \,g ) ( h\, \partial_- g^{-1})\,,
\ee

\no the total effective action can 
be written as,

\be\label{effa}
{\cal S} [U,V,\bar \chi,\chi ] =  S [\bar \chi, \chi, U, V ] 
+ S^\prime [U, V]\,,
\ee

\noindent where

\be\label{c1}
S [\bar \chi, \chi, U, V ] =  S^{(0)}_F[\bar \chi,\chi] + S_M [UV]
\,,
\ee

\be\label{c2}
S^\prime [U, V ] = -\,
\frac{1}{2\,\tilde \mu_o^2}\,\Big \{ \tilde \mu_o^2\,a \,+\,m_o^2\,\Big \}\,
\Gamma [ UV ]\,
+\,\frac{1}{2\,\tilde \mu_o^2}\,\Big \{ \tilde \mu_o^2\,
(a - 2)\,+\,m_o^2\,\Big \}\,\Big (\,\Gamma [V] + \Gamma [U]\,\Big )\,.
\ee

\no The standard TW model corresponds to $a = 2$ and in the 
limit $m_o \rightarrow 0$, the gauge invariance is restored,

\be
S^\prime [U, V ] \rightarrow S^\prime [ {\G} ] \,,
\ee

\no such that the action (\ref{effa}) reduces to the action of the 
standard $QED_2$,

\be
{\cal S} [U,V,\bar \chi,\chi ] \rightarrow {\cal S}_{_{\!QED}} [ {\G},\bar \chi,\chi ]\,.
\ee

\no For a gauge noinvariant regularization, in the 
limit $m_o \rightarrow  0$, we obtain the action for 
the $QED_2$ with broken gauge symmetry,

\be \label{qeda}
S^\prime [U,V,b,0]_{_{\!QED}}\,= -\,
\frac{b}{2}\,\Gamma [ UV ]\,
+\,\frac{1}{2}\,(b - 2)\,\Big (\,\Gamma [V] + \Gamma [U]\,\Big )\,.
\ee

\no where $b$ is the corresponding regularization parameter. The 
actions (\ref{c2}) and (\ref{qeda}) are equivalent,

\be
S^\prime [U,V,b,0] \equiv S^\prime [U,V,a,m_o]\,,
\ee

\no  provided that $b > a$ and

\be\label {bmvf}
m_o^2 = \tilde \mu_o^2 ( b - a )\,.
\ee

\no This implies that the $QED_2$ quantized with a GNI 
regularization $b \neq 2$ ($QED_2$ with broken gauge symmetry) is
isomorphic to the TW model with a regularization parameter $a$ and with
a fixed bare mass for the vector field given by (\ref{bmvf}).

In order to proceed further the decomposition of the Bose field 
algebra $\Im_B \{U,V\}$ into GI and GNI field subalgebras, let us 
introduce the {\it gauge invariant} pseudo-scalar  field $\widetilde{\bphi}$, 

\be
\widetilde{\bphi}\,\doteq\,\frac{1}{2}\,( u\,+\,v )\,,
\ee

\noindent and the {\it  gauge non-invariant} scalar 
field $\zeta$,

\be
\zeta \,\doteq\,\frac{1}{2}\,( u\,-\,v )\,.
\ee

\no The Bose fields $(U,V)$, defined by (\ref{U})-(\ref{V}),  can be 
decomposed in terms of the {\it gauge invariant} 
field ${{\g}} [ \widetilde{\bphi} ]$ and the {\it gauge non-invariant} 
field $h [\zeta ]$ as

\be
U = {\g}\,h\,,
\ee

\be
V = {\g}\,h^{-1}\,,
\ee

\noindent where           

\be
{\g}\,=\,e^{\,\textstyle\,2\,i\,\sqrt \pi\,\widetilde{\bphi}}\,,
\ee

\be
h\,=\,e^{\,\textstyle\,2\,i\,\sqrt \pi\,\zeta}\,.
\ee

\noindent The gauge invariant field ${\G}$ can be rewritten as

\be
{\G} = \,U\, V \,= {\g}\,^2\,=\,
e^{\textstyle\,4\, i\, \sqrt \pi\, \widetilde{\bphi}}\,.
\ee

\noindent The vector  field can be decomposed in the standard form in terms
of GI and  GNI contributions, 
 
\be
{\cal A}_\mu\,=\,\frac{\sqrt \pi}{e}\,\Big \{\,\tilde \partial_\mu\,
\widetilde{\bphi} + \partial_\mu\,\zeta \Big \}\,,
\ee

\noindent and the Maxwell action (\ref{ma}) is now given by 

\be
S_M [ {{\g}} ] \,= \,\frac{1}{2\,\tilde \mu_o^2}\,\int d^2 z\,
(\,\Box\,\widetilde{\bphi}\,)^2\,.
\ee

The Bose field algebra can be decomposed into

\be
\Im^B \{U,V\} = \Im^B_{_{\!{GI}}} \{{\g} \} \oplus \Im^B_{_{\!GNI}} \{ h \}\,,
\ee

\noindent and the effective quantum action ${\cal S} [U, V, \bar \chi, \chi ]$ 
can be rewritten in terms of the fields ${\g}$ and $h$,

\be
{\cal S} [U, V, \bar \chi, \chi ] = {\cal S} [ {\g} , h, \bar \chi, \chi ] =
S [ {\g}, \bar \chi, \chi ] + S^\prime [ {\g}, h ]\,,
\ee

\noindent where

\be
S [ {\g}, \bar \chi, \chi ] = S_F^{(0)} [ \bar \chi, \chi ]  + 
S_M [ {\g} ]\,,
\ee

$$
S^\prime [ {\g}, h ] = \,-\,
\frac{1}{2\,\tilde \mu_o^2}\,\Big \{ \tilde \mu_o^2\,a \,+\,m_o^2\,\Big \}\,
\Gamma [ {{\g}}^2 ] 
$$

\be\label{lgh}
+\,
\frac{1}{2\,\tilde \mu_o^2}\,\Big \{ \tilde \mu_o^2\,
(a - 2)\,+\,m_o^2\,\Big \}\,\Big \{\,\Gamma [ {\g} h^{-1}]\,+\,
\Gamma [{\g} h]\,\Big \}\,,
\ee

\noindent with the Maxwell action given by

\be\label{maxa}
S_M [ {\g} ]\,=\,\frac{1}{8\,\pi\,\tilde \mu_o^2}\,
\int\,d^2 z\,\Big [\,
\partial_+\,(\,{\g}\,i\,\partial_-\,{\g}\,^{-1})\,\Big ]^2\,.
\ee

\noindent Using the P-W identity,  the GNI 
action $S^\prime [ {\g}, h ]$ given by (\ref{lgh}) decouples into, 

\be\label{axy}
S^\prime [ {\g}, h]\,=\,S [ {\g} ] + S [ h ]\,=
\,-\,\frac{\tilde m^2}{\tilde \mu_o^2}\,\Gamma [{\g} ]\,+\,
 \frac{\tilde m_o^2}{\tilde \mu_o^2}\,\Gamma [ h ]\,\,
\ee

\noindent where

\be
\tilde m^2\,=\,\tilde \mu_o^2 ( a + 2 ) + m_o^2\,,
\ee

\be
\tilde m_o^2\,=\,\tilde \mu_o^2 ( a - 2 ) + m_o^2\,.
\ee

\no The fields ${\g} [\tbphi ]$, $h [ \zeta ]$, decouple in the 
action (\ref{axy}) and the GNI
field $h [\zeta ]$ is a free massless non-canonical scalar field. The 
total partition function factorizes as

\be
\Z = \Big ( \int {\cal D} h\,e^{\,i\,S [h]}\,\Big )\,\Big ( \int {\cal D} {\g}\,
{\cal D} \bar \chi\,{\cal D} \chi\,e^{\,i\,S [ {\g} ] + 
S [ {\g}, \bar \chi, \chi ]} \Big )\,=\, {\Z}_h \times 
{\Z}_{\bar \chi, \chi, \g}\,.
\ee

\no The partition function $\Z_{_{h}}$ characterizes the 
local gauge symmetry breakdown.  Although the partition function 
can be factorized, the generating functional, and thus the 
Hilbert space of states ${\cal H}$, cannot be 
factorized

\be
{\cal H} \neq {\cal H}_{_{\!\bar \chi, \chi, {\g}}} \otimes {\cal H}_{_{\!h}}\,.
\ee

The bonafide gauge invariant vector model corresponds 
to $\tilde m_o = 0$, that is obtained with 
$m_o = 0$ and the gauge invariant regularization $a = 2$. In this case, all
reference to the field $h [ \zeta ]$ has desappeared from the effective 
quantum action (\ref{axy}), which
is given in terms of the gauge invariant field ${\g}$. In this case, the 
field $\zeta$ is not a dynamical degree of freedom
and corresponds to a pure $c$-number gauge excitation. The 
commutator (\ref{nccr}) and the corresponding Hamiltonian of the
effective bosonized quantum action are singular for $\tilde m_o = 0$. For these
critical values of the parameters, the GNI action vanishes,

\be
S [h] = \frac{\tilde m_o^2}{\tilde \mu_o^2}\,\Gamma [h] \rightarrow  0\,,
\ee

\noindent implying that for $\tilde m_o = 0$, the gauge invariance
is formally restored and the field \,$ h [\zeta ] $\, is not a 
dynamical degree of freedom. At this critical point the constraint 
structure of the model change. Since in the gauge invariant 
limit $\tilde m_o \rightarrow 0$ the field $h$ becomes a pure gauge 
excitation, the corresponding partition function reduces to a gauge volume,

\be
\lim_{\tilde m_o \rightarrow 0}\,\Z_h \,
\rightarrow \int {\cal D}\,h = V_{_{gauge}}\,.
\ee

The ``anomalous \textbf{vector} Schwinger model'' considered in Ref. \cite{Tiao} is obtained
considering $m_o = 0$ and $b \neq 2$. The GNI action is now given by

\be
S^\prime [{\g},h]\, =\, -\,( b + 2 )\,\Gamma [{\g}]\,+\,
( b - 2 ) \,\Gamma [h]\,,
\ee

\no and corresponds to the TW model with bare mass for the vector field given
by (\ref{bmvf}).

%$\#\#\#\#\#\#\#\#\#\#\#\#\#\#\#\#\#\#\#\#$
%%%%%%%%%%%%%%%%%%%%%%%%%%%%%%%%%%%
\subsection{Local Action}
%%%%%%%%%%%%%%%%%%%%%%%%%%%%%%%%%%

The Maxwell action (\ref{maxa}) is non-local due to the 
quartic-self-interaction of the field ${\g}$. In 
order to dequartize the action of the gauge invariant field ${\g}$, let 
us consider the functional integral over the field ${\g}$ in the partition 
function ${\Z}_{\bar \chi, \chi, \g}$. To begin with, let us introduce 
an auxiliary gauge invariant field $\opOmega$, such that,

$$
\int {\cal D} {\g}\, \exp \,i\,\int\,d ^2 z\,\Big \{\,
\,\frac{1}{8 \pi \tilde \mu_o^2}\, [\,
\partial_+\,(\,{\g}\,i\,\partial_-\,
{\g}\,{^{- 1}}\,)\, ]^2\,
-\,\frac{1}{8 \pi}\,\frac{\tilde m^2}{\tilde \mu_o^2}\,
\,\partial_+ {\g}\,\partial_- {\g}^{- 1} 
\Big \}\,\equiv
$$

$$
\int\,{\cal D} \opOmega\,{\cal D} {\g}\,\exp
\,i\,\int\,d ^2 z\,
\Big \{\,-\,\frac{1}{2}\,\opOmega^2\,+\,\frac{1}{2 \sqrt \pi \tilde \mu_o}\,
\opOmega\,[\,\partial_-\,(\,{\g}^{- 1}\,i\,\partial_+\,
{\g}\,)\, ]
$$

\be \label{dequart}
-\,\frac{1}{8 \pi}\,\frac{\tilde m^2}{\tilde \mu_o^2}
\partial_+ {\g}\,\partial_- {\g}^{- 1}\,\Big \}\,.
\ee

\noindent Making the change of variables,

\be
\partial_- \opOmega\,\doteq\,(\,\opomega\,i\,\partial_- \opomega ^{-1})\,,
\ee

\noindent we can writte,

\be
\opOmega = \,\partial_-^{-1}\,
(\,\opomega\,i\,\partial_- \opomega ^{-1})\,.
\ee

\noindent Rescaling the exponential field $\opomega$,

\be
\frac{2 \sqrt \pi \tilde \mu_o}{\tilde m^2}\,\opomega^{-1}\,i\,
\partial_+\,\opomega\,=\,{\check{\opomega}}^{-1}\,i\,
\partial_+\,{\check{\opomega}}\,,
\ee

\noindent we obtain from (\ref{dequart}) after an integration by parts,

$$
\int\,{\cal D} \copomega\,{\cal D} {\g}\,
\exp \,i\,\int\,d ^2 z\,
\Big \{\,-\,\frac{1}{8 \pi}\,\frac{\tilde m^4}{\tilde \mu_o^2}\,
[\,\partial_-^{-1} (\copomega \,i\,\partial_-\,\copomega^{-1})\,]^2\,
+ 
$$

$$
+\,\frac{1}{8 \pi}\,\frac{\tilde m^2}{\tilde \mu_o^2}\,\Big (\,
{\g}\,^{-1}\,i\partial_+ {\g}\,-\,\copomega^{-1} \,i\,
\partial_+\,\copomega \,\Big )\,
\Big (\,{\g}\,i\partial_-\, {\g}\,^{-1}\,-\,
\copomega \,i\,\partial_-\,\copomega^{-1} \,\Big )
$$

\be
-\,\frac{1}{8 \pi}\,\frac{\tilde m^2}{\tilde \mu_o^2}\,
(\copomega^{-1}\,i\,\partial_+\,\copomega\,)\,
(\,\copomega\,i\,\partial_-\,\copomega^{-1}\,)\,.
\ee

\noindent Defining a new gauge invariant field $\optheta$,

\be
\optheta^{-1}\,i\,\partial_+\,\optheta\,\doteq\,
{\g}\,^{-1}\,i\partial_+ 
{\g}\,-\,\copomega ^{-1} \,i\,\partial_+\,\copomega \,,
\ee

\be
\optheta\,i\,\partial_-\,\optheta^{-1}\,\doteq\,
{\g}\,i\partial_-\,{\g}\,^{-1}\,-\,\copomega \,i\,
\partial_-\,\copomega^{-1}\,,
\ee

\noindent the field ${\g}$ can be factorized as,

\be
{\g} = \optheta\,\copomega\,.
\ee

\noindent The total effective action is now written in terms of the 
gauge invariant fields $\optheta$, $\copomega$, the gauge noninvariant field 
$h$ and the free Fermi field $\chi$. In order to obtain the complete bosonized action, we introduce the 
bosonized form for the action of the free Fermi field \cite{AAR},

\be
S_F^{(0)} = \Gamma [ f_{_{\!\varphi}} ] = \frac{1}{2}\,
\int d^2 z\,\partial_\mu \varphi\,\partial^\mu \varphi\,,
\ee

\noindent with

\be
f_{_{\!\varphi}} = e^{\,2\,i\,\sqrt \pi \,\varphi}\,,
\ee

\noindent and the Mandelstam representation for the bosonized free massive
Fermi field,

\be
\chi (x) = \Big ( \frac{\kappa_o}{2 \pi} \Big )^{\frac{1}{2}}\,
e^{\,-\,i\,\frac{\pi}{4}\gamma^5}\,
: exp \,i\,\sqrt \pi\,\Big \{\,\gamma^5\,\tilde \varphi (x)\,
\, + \,\int_{x^1}^\infty\,dz^1\,\partial_0\,\tilde \varphi (x^0,z^1)\,
\Big \} : \,,
\ee

\noindent where $\kappa_o$ is an arbitrary finite mass scale. The total 
local action is given in the bosonized form by,

$$
{\cal S} [ f, {\g}, h ]\,=\,{\cal S} [ f, \optheta, \copomega, h ]\,=\,
 \Gamma [f_{_{\!\varphi}}]\,-\,\Gamma [ \optheta ]\,+\,
\Gamma [ \copomega  ]\, +\,
\frac{\tilde m_o^2}{\tilde \mu_o^2}\,\Gamma [ h ]\,-
$$

\be
-\,\,\frac{1}{8 \pi}\,\frac{\tilde m^4}{\tilde \mu_o^2}\,\int d^2z\,
[\,\partial_-^{-1}\,(\copomega \,i\,\partial_-\,
\copomega  ) ]^2\,,
\ee

\noindent where the field $\optheta$ is quantized with 
negative metric. Parametrizing the fields $\optheta$ and $\copomega$ as, 

\be
\optheta (x)\,=\,e^{\textstyle\,2\,i\,\sqrt \pi \, \tilde \eta (x)}
\,,\ee

\be
\copomega (x)\,=\,e^{\textstyle\,2\,i\,\sqrt \pi \,
\frac{\tilde \mu_o}{\tilde m^2}\, \tilde \Sigma (x)}\,,
\ee

\noindent  performing the field scaling,

\be
\tilde \eta \rightarrow \frac{\mu_o}{\tilde m}\,\tilde \eta^\prime\,,
\ee

\be
\tilde \Sigma \rightarrow \tilde m\,\tilde \Sigma^\prime\,,
\ee

\noindent and streamlining the notation by dropping primes everywhere, we 
obtain for the gauge invariant field $\tbphi$,

\be
\tbphi = \frac{\tilde \mu_o}{\tilde m}\,(\tilde \eta + \tilde \Sigma ) 
\,,
\ee

\be
{\g}   = \optheta\,\copomega = e^{\textstyle\,2 i \sqrt \pi\,\frac{\tilde \mu_o}{\tilde m}\,
(\tilde \eta + \tilde \Sigma )}\,.
\ee

\no The  effective bosonized total Lagrangian density is then given by,

\be\label{ebtld}
{\cal L} =   \frac{1}{2} \,(\partial_\mu  \tilde \varphi)^2 +
\frac{1}{2}\,(\partial_\mu \tilde \Sigma )^2 - \,\frac{\tilde m^2}{2}\, 
\tilde \Sigma^2 - \frac{1}{2}\,(\partial_\mu \tilde \eta )^2\,+ \,\frac{1}{2}\,
\frac{\tilde m_o^2}{\tilde \mu_o^2}\,(\partial_\mu \zeta )^2 \,.
\ee

\noindent The field $\tilde \eta$ is quantized with negative metric. The 
gauge non-invariant field $\zeta$ is a non-canonical free massless 
decoupled field,

\be\label{nccr}
[ \zeta (x)\,,\,\zeta (y) ]\,=\,\frac{\tilde \mu_o^2}{\tilde m_o^2}\,\Delta (x - y;0)\,.
\ee

\noindent For massless Fermi fields, although the total partition function 
factorizes into free field partition functions,

\be
\Z\,=\,\Z_{_{f}}\,\times\,\Z_{_{\theta}}\,
\times\,\Z_{_{\omega}}\, 
\times\, \Z_{_{h}}\,,
\ee

\no the generating functional does not factorizes.

%%%%%%%%%%%%%%%%%%%%%%%%%%%%%%%%
\section{Field Operators and Hilbert Space}
%%%%%%%%%%%%%%%%%%%%%%%%%%%%%%%
\setcounter{equation}{0}

The Bose fields $(U,V)$ are given by,

\be\label{opu}
U = e^{\,2\,i\,\sqrt \pi\,\tilde{\bphi} }\,h [\zeta ] =
e^{\,2\,i\,\sqrt \pi\,\frac{\tilde \mu_o}{\tilde m} 
( \tilde \eta + \tilde \Sigma )}\,h [ \zeta ]\,,
\ee

\be\label{opv}
V = e^{\,2\,i\,\sqrt \pi\,\tilde{\bphi} }\,h^{-1} [\zeta ] =
e^{\,2\,i\,\sqrt \pi\,\frac{\tilde \mu_o}{\tilde m} 
( \tilde \eta + \tilde \Sigma )}\,h^{-1} [ \zeta ]\,.
\ee

\noindent The bosonized form of the Fermi field operator is given in terms of
the free Fermi field as,

$$
\psi (x) = \Big (\frac{\kappa_o}{2 \pi} \Big )\,e^{- i \frac{\pi}{4}\gamma^5}\,
:e^{\,2\,i\,\sqrt \pi\,\frac{\tilde \mu_o}{\tilde m}\,\gamma^5\,\{\tilde \eta (x) +
\tilde \Sigma (x)\}}: \times
$$

\be\label{ff}
:e^{\,\,i\,\sqrt \pi\,\{\,\gamma^5\,\tilde \varphi (x) +
\int_{x^1}^{\infty} \partial_0\,\tilde \varphi (x^0,z^1)\,dz^1\}}:\,h^{-1} [\zeta ],
\ee

\no and the vector field is given by, 

\be\label{gauf}
{\cal A}_\mu\,=\,
\frac{1}{\tilde m}\,\tilde \partial_\mu\,( \tilde \Sigma + \tilde \eta )\,+\,
\frac{1}{e}\,h\,i\,\partial_\mu\,h^{-1}\,.
\ee

\no The divergent part $h [ \zeta ]$ has the form of a gauge term. The 
bosonized expressions for fields (\ref{ff}) and (\ref{gauf}) correspond to 
those of the Schwinger model gauged by a divergent operator-gauge 
transformation as long as $\tilde m_o \rightarrow 0$. 

The vector current is computed with the standard point-splitting 
limit procedure,

\be\label{ps}
J_\mu (x)\,= \vdots \bar \psi (x)\,\gamma_\mu\,\psi (x)\vdots\, =\,Z^{-1} (\epsilon )\,\Big [ \bar \psi (x + \epsilon )\,\gamma_\mu\,
e^{\,i\,a\,\tilde \mu_o\,\int_x^{x + \epsilon }\,{\cal A}_\mu (z) dz^\mu}\,
\psi (x)\,-\,V.E.V. \Big ]\,.
\ee

\no In terms of the GI and GNI field combinations $ (u \pm v) $, the vector 
current is given by,

\be\label{vcur1}
\frac{e}{2}\,J_\mu \,=\,\frac{e}{2}\,\j_\mu\,-\,\frac{\tilde \mu_o}{2}\,
( a + 2 )\,\tilde \partial_\mu ( u + v )\,
-\,\frac{\tilde \mu_o}{2}\,( a - 2 )\,
\partial_\mu ( u - v )\,,
\ee

\no where  $\j_\mu$ is the conserved current associated with the free 
Fermi field $\chi$,

\be
\j_\mu = \vdots \bar \chi \gamma_\mu \chi \vdots = -\,\frac{2}{\sqrt \pi}\,\tilde \partial_\mu 
\tilde \varphi\,.
\ee

\no The bosonized form of the vector current is given by,

\be\label{vcur2}
\frac{e}{2}\,J_\mu \,=\,-\,2\,\tilde \mu_o\,\tilde \partial_\mu \tilde
\varphi\, +\,\frac{(\tilde m - m_o^2)}{\tilde m}\,
\tilde \partial_\mu ( \tilde \Sigma + \tilde \eta )\,
-\,\tilde \mu_o\,( \tilde m_o^2 - m_o^2 )\,
 \partial_\mu\,\zeta\,.
\ee

\no The conserved current that acts as the source of ${\cal F}_{\mu \nu}$ is given
by

\be
{\cal K}_\mu\,=\,m_o^2\,{\cal A}_\mu\,-\,\frac{e}{2}\,J_\mu\,=\,
-\,\frac{e}{2}\,\j_\mu\,+\,\frac{\tilde m^2}{2 \tilde \mu_o}\,
\tilde \partial_\mu ( u + v )\,
+\,\frac{\tilde m_o^2}{2 \tilde \mu_o}\,\partial_\mu ( u - v )\,,
\ee

\noindent which can be written as,  

\be\label{veccur}
{\cal K}_\mu \,=\,\tilde m\,\tilde \partial_\mu\,\tilde \Sigma\,+\,L_\mu\,,
\ee

\noindent where $L_\mu$ is a longitudinal current of zero norm given by,

\be\label{cur2}
L_\mu\, =\,\partial_\mu\,L\,,
\ee

\no with the potential $L$ given by,

\be
L\,=\,  \,2\,\tilde \mu_o\,\varphi\,+\,\tilde m\,\eta\,+\,
\frac{\tilde m_o^2}{\tilde \mu_o}\,\zeta\,.
\ee

\no The longitudinal current $L_\mu$ commutes with itself and thus generates
zero norm states from the vacuum,

\be
\langle \Psi_o \vert L_\mu (x) L_\nu (y) \vert \Psi_o \rangle = 0\,.
\ee

\no Due to the presence of the longitudinal current $L_\mu$, the Proca equation
is not satisfied as an operator identity,

\be
\partial_\nu\,{\cal F}^{\nu \mu} + m_o^2\,{{\cal A}}^\mu\,-\,
\frac{e}{2}\,{J}^\mu\,=\,L_\mu\,.
\ee

The fundamental fields $\{\bar \psi, \psi, {\cal A}_\mu \}$ generate a local 
field algebra $\Im$. These field operators constitute the intrinsic mathematical
description of the model and whose Wightman functions define the model. The 
field algebra $\Im$ is represented in the indefinite-metric Hilbert space 
of states ${\cal H}$,

\be
{\cal H} = \Im \vert \Psi_o \rangle\,.
\ee

In a {\it gauge invariant model}, the field $\zeta$
is not a dynamical degree of freedom. In the indefinite metric 
formulation, the Gauss' law is satisfied in weak form,

\be
\partial_\nu {\cal F}^{\nu \mu}\,-\,e\,J^\mu = L^\mu\,,
\ee

\noindent where $L_\mu$ is the zero-norm longitudinal piece of the vector 
current that acts as the source in the Gauss' law. The 
{\it local gauge transformations} of the intrinsic 
fields $\{\bar \psi, \psi, {\cal A}_\mu \}$ are implemented by the 
longitudinal current $L_\mu$. The {\it physical} gauge invariant 
field algebra $\Im_{_{phys}}$, is a subalgebra of the intrinsic field algebra 
$\Im$ that commutes with the longitudinal current,

\be
\Im_{_{phys}} \subset \Im\,,
\ee
\be
[ \Im_{_{phys}}\,,\,L_\mu ] = 0\,.
\ee

\noindent The physical Hilbert 
space 

\be
{\cal H}_{_{phys}} = \Im_{_{phys}} \vert \Psi_o \rangle\,,
\ee

\no is a subspace of the Hilbert space ${\cal H} = \Im \vert \Psi_o \rangle$,

\be
{\cal H}_{_{phys}} \subset {\cal H}\,.
\ee

In a model with {\it gauge symmetry breakdown}, the field $\zeta$ is a dynamical 
degree of freedom. {\it The intrinsic field algebra commutes with the 
longitudinal current},

\be
[ \Im , L_\mu ] = 0\,.
\ee

\no This implies that the fields $\{ \bar \psi, \psi, {\cal A}_\mu\}$ are 
singlet under gauge transformations generated by the longitudinal current 
and thus are {\it physical operators}. {\it The intrinsic field algebra is by 
itself the physical field algebra},

\be
\Im \equiv \Im_{_{\!phys.}}\,.
\ee

%%%%%%%%%%%%%%%%%%%%%%%%%%%%%%%
\subsection{The $QED_2$ limit}
%%%%%%%%%%%%%%%%%%%%%%%%%%%%%

It is very instructive to make some comments about the gauge invariant 
limit $\tilde m_o \rightarrow 0$. From the Lagrangian (\ref{ebtld}), written 
in terms of the non-canonical field $\zeta$, we obtain in the zero mass
limit,

\be
{\cal L}_{_{\tilde m_o \rightarrow 0}}\,\rightarrow\,
{\cal L}_{_{QED}}\,.
\ee

\no In this indefinite-metric Hilbert space formulation, the GI limit can be 
performed in the operator field algebra written in terms of the non-canonical
free field $\zeta$. The operator solution of the $QED_2$, as  given
by Lowenstein-Swieca \cite{LS}, can be formally 
obtained from (\ref{ff}),(\ref{gauf}), (\ref{veccur}) and (\ref{cur2}). In the
gauge invariant limit, the field $\zeta$ decouples from the quantum action 
corresponding to the Lagrangian (\ref{ebtld}) and thus it is 
not a dynamical degree of freedom. The field $\zeta$ becomes a 
$c$-number, $\zeta (x) \rightarrow \Lambda (x)$, and we get,

\be
\psi (x) = :e^{\,i\,\sqrt \pi\,\gamma^5\,\{\tilde \eta (x) +
\tilde \Sigma (x)\}}:\,\chi (x)\,e^{\,-\,i\,\sqrt \pi \Lambda (x)}\,,
\ee

\be
{\cal A}_\mu\,=\,\frac{\sqrt \pi}{e}\,\tilde \partial_\mu\,
( \tilde \Sigma + \tilde \eta )\,+\,\frac{\sqrt \pi}{e}\,\partial_\mu\,
\Lambda (x)\,,
\ee

\be
J_\mu = \frac{1}{\sqrt \pi}\,\tilde \partial_\mu \tilde \Sigma\,+\,L_\mu\,,
\ee

\noindent where $L_\mu$ is the longitudinal current of zero norm,

\be
L_\mu = - \frac{e}{\sqrt \pi}\,\partial_\mu\,( \varphi + \eta )\,.
\ee

\no Taking this into account, we obtain for the generating functional,

\be
{\cal Z} [ \bar \vartheta, \vartheta, \j_\mu ]_{_{\tilde m_o \rightarrow 0}}\,
\rightarrow\,{\cal Z} [ \bar \vartheta, \vartheta, \j_\mu ]_{_{QED}}\,.
\ee

\no It must be stressed that this gauge invariant limit can be formally 
performed only in this level. It cannot be performed in the general 
Wightman functions due to the singular commutation relation for the 
field $\zeta$. This limit is well defined
only for the subset of Wightman functions that are 
independent of the field $\zeta$. By considering the quantum action and the
field operators written in terms of the non-canonical field $\zeta$, to take the 
limit $m_o \rightarrow 0$ is equivalent to start from the beginning 
with a zero bare mass for the vector meson. As a matter of fact, in this 
indefinite-metric Hilbert space
formulation, the GI limit can be formally performed as above, since 
the Fermi field operator is written in terms of the free 
Fermi field and not in terms of the charge-carrying Fermi field operator 
of the Thirring model. The intrinsic field algebra $\Im$ is the physical 
field algebra, which is singlet under gauge transformations generated by the
longitudinal current.

%%%%%%%%%%%%%%%%%%%%%%%%%%%%%%%%%%%%%%%%
\section{The Positive-definite-metric Formulation}
%%%%%%%%%%%%%%%%%%%%%%%%%%%%%%%%%%%%%%%%%
\setcounter{equation}{0}

The bosonization in the indefinite-metric formulation introduces a larger 
Bose field algebra $\Im_B$  that 
contains more degrees of freedom than those needed for the description 
of the model, $\Im \subset \Im_B$. In order to extract the redundant degrees of 
freedom, as well 
as, to obtain the solution of the equations of motion as operator identities
in a positive definite-metric Hilbert space of states, we introduce 
two free Bose fields $(\tilde \Phi,\tilde \Xi)$, by the following canonical 
transformation

\be\label{ct1}
\frac{\beta}{2}\,\,\tilde \Phi\,=\,\sqrt \pi\,\tilde \varphi\,+\,2\,\sqrt \pi\,
\frac{\tilde \mu_o}{\tilde m}\,\tilde \eta\,,
\ee

\be\label{ct2}
\frac{\beta}{2}\,\tilde \Xi\,=\,\,2\,\sqrt \pi\,\frac{\tilde \mu_o}{\tilde m}\,
\tilde \varphi\,+\,\sqrt \pi\,\tilde \eta\,.
\ee

\noindent The negative metric quantization for the field $\tilde \eta$
ensures that the fields $\tilde \Phi$ and $\tilde \Xi$ are independent degrees
of freedom,

\be
[ \tilde \Phi (x)\,,\,\tilde \Xi (y) ]\,=\,0\,\,\,,\,\,\,\forall\,(x,y)\,.
\ee

\noindent Imposing canonical commutation relations for the fields $\Phi$ and 
$\Xi$, we get

\be
\frac{\beta^2}{4 \pi}\,=\,\Big (\,\frac{\tilde m_o^2}{\tilde m^2}\,\Big )\,>\,0\,,
\ee

\noindent and the field $\tilde \Xi$ is quantized with negative 
metric. Definig the canonical free field,

\be
\xi = \frac{\tilde m_o}{\tilde \mu_o}\,\zeta\,,
\ee

\no the Fermi field and the vector field operators (\ref{ff}), (\ref{gauf}) 
can be re-written as `` gauge transformed fields'',

\be\label{ptw}
{\psi}\,=\,\widehat \psi\,{\rho}\,,
\ee

\be\label{atw}
{\cal A}_\mu\,=\,\widehat {\cal A}_\mu\,+\,\frac{1}{e}\,\rho\,i\,\partial_\mu\,\rho^{-1}\,.
\ee

\noindent Here, the operator $\rho$ is a pure gauge excitation given by,

\be
\rho\,=\,:e^{\textstyle \,2\,i\,\sqrt \pi\,
\frac{\tilde \mu_o}{\tilde m_o}\,(\Xi - \xi)}:\,,
\ee

\no and 

\be
\widehat \psi \,=\,:e^{\textstyle \,-\,2\,i\,\gamma^5\,
\sqrt \pi\,\frac{\tilde \mu_o}{\tilde m}\,\tilde \Sigma }:\,
{\Psi}_{_{\tilde \Phi}}\,,
\ee

\be
\widehat A_\mu \,=\,-\,\frac{1}{\tilde m}\,\tilde \partial_\mu\,\Big (\,\tilde \Sigma\,
-\,2\,\frac{\tilde \mu_o}{\tilde m_o}\,\tilde \Phi\,\Big )\,,
\ee

\noindent Here $\Psi$ is the charge-carrying Fermi field operator of the Thirring model, which is 
given by the Mandelstam representation,
 
\be
{\Psi} (x)\,=\,\Big (\,\frac{\mu_o}{2 \pi} \Big )^{\frac{1}{2}}\,e^{- i \frac{\pi}{4}\gamma^5}\,
:\exp\, \Big \{\,-\,i\,
\,\gamma^5\,\frac{\beta}{2}\,\tilde \Phi (x)\,-\,i\,
\frac{2\pi}{\beta}\,\int_{x^1}^{+ \infty} \partial_0\,{\tilde \Phi} (x^0,z^1) dz^1 \Big \}
:\,.
\ee

\no The vector current ${J}_\mu$, given by (\ref{veccur}), can 
be re-written as

\be
\frac{e}{2}\,{J}_\mu\,=\,
\frac{(\tilde m^2 -  m_o^2)}{\tilde m}\,
\tilde \partial_\mu \tilde \Sigma\,+\,\frac{2 \tilde \mu_o\,m_o^2}{\tilde m_o \tilde m}\,
\tilde \partial_\mu \tilde \Phi\,.
\ee

\no and the current ${\cal K}_\mu$ is given by,

\be
{\cal K}_\mu\,=\,\tilde m\,\tilde \partial_\mu\,\tilde \Sigma\,+\,L_\mu\,,
\ee

\no where the longitudinal current is now given by,

\be
\ell_\mu = \partial_\mu L = 
\tilde m_o \partial_\mu \,( \Xi - \xi )\,=\,\frac{\tilde m_o}{2 \sqrt \pi\,\tilde
\mu_o}\,\rho\,\partial_\mu\,\rho^{-1}\,.
\ee

\no The longitudinal current carries no fermion selection rule since it
is independent of the vector current of the Thirring model,

\be
J_\mu^{Th} = \frac{2 \tilde \mu_o m_o^2}{\tilde m_o \tilde m}\,
\tilde \partial_\mu \tilde \Phi\,.
\ee

For the GI regularization $a = 2$, we obtain,

\be
\frac{\beta^2}{4 \pi} \,= \,\frac{m_o^2}{\frac{e^2}{\pi} + m_o^2}\,,
\ee

\be
J_\mu^{Th} = \frac{\beta}{\sqrt \pi}\,
\tilde \partial_\mu \tilde \Phi\,,
\ee

\no and the field operators given by (\ref{ptw})-(\ref{atw}) correspond
to those operators obtained by Lowenstein-Rothe-Swieca \cite{LS,RS}
gauged by a singular operator gauge transformation.

The operator  $\rho$ has zero scale dimension, commutes with itself and 
thus generates infinitely delocalized states that leads to constant 
Wightman functions

\be
\langle \Psi_o \vert\,\rho^\ast (x_1)\,\cdots\,\rho^\ast (x_n)\,\rho (y_1)\,
\cdots\,\rho (y_n)\,\vert \Psi_o \rangle\,=\,1\,.
\ee

\no The spurious operator $\rho$ does not carrie any fermionic charge 
selection rule, and since it  commutes with all 
operators belonging to the
field algebra $\Im$, it is reduced  to the identity operator 
in ${\cal H}$. The position independence 
of the state $\rho (x) \Psi_o$ can be seen by computing the general Wightman functions involving
the operator $\rho$ and all operators belonging to the local field 
algebra $\Im$. Thus, for any operator ${\cal O} (f_z) = \int {\cal O} (z)\,
f (z)\,d\,^2 z\,\in\,\Im$, of 
polynomials in the smeared 
fields $\{\bar \psi, \psi, {\cal A}_\mu \}$, the position independence  of the
operator $\rho$ can be expressed in the weak form as

$$
\langle \Psi_o\,,\,\rho^\ast (x_1)\cdots \rho^\ast (x_\ell)\rho(y_1)\cdots
\rho (y_m)\,{\cal O} (f_{z_1}, \cdots, f_{z_n})\,\Psi_o \rangle\,=
$$

\be
{\cal W} (z_1,\cdots,z_n)\,\equiv\,
\langle \Psi_o\,, {\cal O} (f_{z_1}, \cdots, f_{z_n})\,\Psi_o \rangle\,
\,,\,\,\forall\,{\cal O} (f_{z})\,\in\,\Im\,,
\ee

\noindent where ${\cal W} (z_1,\cdots,z_n)$ is a distribution independent of 
the space-time coordinates $(x_i,y_j)$.

Within the functional integral approach, the Hilbert space of states is 
constructed from  the generating functional (\ref{gf}), which can be rewritten
as,  

\be\label{bgfunct}
{\Z}\,[\,\bar \vartheta, \vartheta, \jmath^\mu\,]\,=\,
\Big \langle\,
\exp \,i\,\int\,d^2 z\,\Big (\,\bar \vartheta \,\widehat \psi \,\rho\,+\,
\bar {\widehat \psi}\,\rho^\ast\,\vartheta\,+\,\jmath^\mu\,( \widehat A_\mu \,+\,
\frac{1}{e}\,\rho^{-1}\,i\,\partial_\mu\,\rho )\Big ) \Big \rangle\,,
\ee

\noindent where the average is taken with respect to the 
functional integral measure,

\be\label{bmes}
\mes\,=\,
\int\, {\cal D}\,\Xi\, {\cal D}\,\xi\,e^{\,i\,S^{(0)} [ \Xi, \xi ]}\,
\int\,{\cal D}\,\tilde \Phi\,{\cal D}\,\tilde \Sigma\,
e^{\textstyle \,i\,{\cal S} [\tilde \Sigma\,\tilde \Phi ]}\,,
\ee

\no and the actions appearing in (\ref{bmes}) are  given in terms of the 
corresponding bosonized Lagrangian densities,

\be
{\cal L}^{(0)} =  - \frac{1}{2}\,(\partial_\mu \Xi )^2\,+ \,\frac{1}{2}\,
\,(\partial_\mu \xi )^2 \,,
\ee

\be\label{LRS}
{\cal L} =   \frac{1}{2} \,(\partial_\mu  \tilde \Phi)^2 +
\frac{1}{2}\,(\partial_\mu \tilde \Sigma )^2 - \,\frac{\tilde m^2}{2}\, 
\tilde \Sigma^2\,.
\ee

\no Since the free massless fields $\Xi$ and $\xi$ decouples 
in the quantum action,  the partition function factorizes,

\be\label{pff}
\Z = \Z^{(0)}_{_{\!\xi}}\times \Z^{(0)}_{_{\! \Xi}} \times \Z^{(0)}_{_{\!\Phi}}
\times \Z^{(0)}_{_{\!\Sigma}}\,.
\ee

\no In the computation of general correlation functions from the generating
functional (\ref{bgfunct}), due to the 
opposite metric quantization, the space-time contributions
of the functional integration over the field $\xi$ cancel those contributions
comming from the integration over the field $\Xi$. In this way, the field
$\rho$ becomes the identity with respect to the functional integration and 
we get the identities,

\be\label{opsol}
\psi \equiv \widehat \psi \,\,\,,\,\,\,{\cal A}_\mu \equiv \widehat {\cal A}_\mu\,.
\ee

\no The positive-definite metric Hilbert space $\widehat {\cal H}$ is builded 
from the generating functional,

\be
{\Z}\,[\,\bar \vartheta, \vartheta, \jmath^\mu\,]\,\equiv\,\widehat {\Z}\,[\,\bar \vartheta, \vartheta, \jmath^\mu\,]\,=\,
\Big \langle\,
\exp \,i\,\int\,d^2 z\,\Big (\,\bar \vartheta \,\widehat \psi \,+\,
\bar {\widehat \psi}\,\vartheta\,+\,\jmath^\mu\, \widehat A_\mu \,\Big ) \Big \rangle\,,
\ee

\no where the average is taken with respect to the functional integral measure

\be
\widehat{\mes}\,=\,
\int\,{\cal D}\,\tilde \Phi\,{\cal D}\,\tilde \Sigma\,
e^{\textstyle \,i\,{\cal S} [\tilde \Sigma\,\tilde \Phi ]}\,.
\ee

\no The Hilbert space $\widehat {\cal H}$ corresponds to the quotient space

\be
\widehat {\cal H} = \frac{{\cal H}}{{\cal H}_0}\,,
\ee

\no where ${\cal H}_0$ is the zero-norm space. The 
fields (\ref{ptw})-(\ref{atw}) provide the 
operator solution for the coupled Dirac-Proca equations

\be
i\,\gamma^\mu \partial_\mu {\psi} (x)\,+\,e\,\gamma^\mu :
{{\cal A}}_\mu (x)\,{\psi}(x):\,=\,0\,,
\ee

\be
\partial_\nu\,{\cal F}^{\nu \mu} + m_o^2\,{{\cal A}}^\mu\,-\,
\frac{e}{2}\,{J}^\mu\,=\,0\,.
\ee

\no For the gauge invariant regularization $a = 2$ and $m_o \neq 0$, we 
recover  from (\ref{opsol}) the operator solution 
for the TW model obtained by Lowenstein-Rothe-Swieca \cite{LS,RS}. For $m_o = 0$ and
the gauge non-invariant regularization $a \neq 2$, we obtain 
the operator solution of $QED_2$ with broken gauge symmetry.

In the positive-metric Hilbert space formulation, the $QED_2$ limit 
does not exist for the Fermi field and
vector field themselves. In this case, the equations of motion are satisfied as operator 
identities and the Fermi field is a charge-carrying operator,

\be
[ J_{_{\!TW}}^0 (x)\,,\,\Psi (z) ]_{_{x^0 = z^0}}\,=\,-\,e\,\frac{m_o^2}{\tilde m_o^2}\,\delta (x^1 - z^1)\,\Psi (z)\,.
\ee

\no As stressed in Ref. \cite{RS}, if this limit were to exist, we 
would obtain for the $QED_2$ a local charge-carrying Fermi field operator, which is 
incompatible with the Maxwell's equation being 
satisfied in the strong form. Nevertheless, this limit
is well defined for the gauge invariant field subalgebra, as for instance,

\be
J_\mu\,\rightarrow\,\frac{1}{\sqrt \pi}\,\tilde \partial_\mu\,\tilde \Sigma\,,
\ee

\be
{\cal F}_{\mu \nu} \rightarrow  \epsilon_{\mu \nu}\,\frac{e}{\sqrt \pi}\,\tilde \Sigma\,.
\ee

%%%%%%%%%%%%%%%%%%%%%%%%%%%%%
\subsection{Massive Fermions}
%%%%%%%%%%%%%%%%%%%%%%%%%%%

The introduction of a mass term for the Fermi field gives a contribution to
the action,

$$
{\cal M}\,=\,- M_o \,\bar \psi\,\psi = 
-\,
M_o\,\Big \{\,\chi_1^\dagger \chi_2\,( U\,V)\,+\,
\chi_2^\dagger \chi_1\,(U\,V )^{-1}\, \Big \}\,=
$$

\be
-\,M_o\,\Big \{\,\chi_1^\dagger \chi_2\,{\g} ^2\,+\,
\chi_2^\dagger  \chi_1\,{\g}^{-2}\, \Big \}\,.
\ee

\no Using the decomposition for the GI field ${\g}$ and the bosonized form for
the free Fermi field and  the 
bosonized chiral density of the free Fermi field,

\be
\chi_1^\ast \chi_2 = \Big (\frac{\kappa_o}{2 \pi}\Big )\,:e^{\,2\,i\,\sqrt \pi\,\varphi}:\,,
\ee

\no we get,

$$
{\cal M}\,=\,\frac{M_o\,\kappa_o}{2 \pi}\,\Big (\,e^{\,2\,i\,\sqrt \pi\,\tilde \varphi}\,
[\copomega \,\optheta ]^2\,+\,
e^{\,-\,2\,i\,\sqrt \pi\,\tilde \varphi}\,[\copomega\,\optheta ]^{-2}\,\Big )
\Big \}\,=
$$

\be 
-\,\frac{M_o\,\kappa_o}{\pi}\,\cos \{\,2\,\sqrt \pi\,\tilde \varphi\,+\,
4\,\sqrt \pi\,\frac{\tilde \mu_o}{\tilde m}\,(\,\tilde \eta\,+\,\tilde \Sigma\,
)\,\}\,.
\ee

\no Performing the canonical transformation (\ref{ct1})-(\ref{ct2}), the mass term can be written as

\be
{\cal M}\,=\,
 -\,\frac{M_o\,\kappa_o}{\pi}\,\cos \{\,2\,\beta \tilde \Phi\,
+\,
4\,\sqrt \pi\,\frac{\tilde \mu_o}{\tilde m}\,\tilde \Sigma\,\}\,.
\ee

\no Notice that, even for a massive fermion field, the 
field $\zeta$ ($\xi$) is a free field. The fields $\tilde \Phi$ and $\tilde \Sigma$ are coupled by the 
sine-Gordon interaction and the total Lagrangian density is now given by

$$
{\cal L} =  - \frac{1}{2}\,(\partial_\mu \Xi )^2\,+ \,\frac{1}{2}\,
\,(\partial_\mu \xi )^2 \,
$$

\be\label{mlag}
+\,\frac{1}{2} \,(\partial_\mu  \tilde \Phi)^2 +
\frac{1}{2}\,(\partial_\mu \tilde \Sigma )^2 - \,\frac{\tilde m^2}{2}\, 
\tilde \Sigma^2\,
 -\,\frac{M_o\,\kappa_o}{\pi}\,\cos \{\,2\,\beta \tilde \Phi\,
+\,
4\,\sqrt \pi\,\frac{\tilde \mu_o}{\tilde m}\,\tilde \Sigma\,\}\,.
\ee

\no For $a = 2$ we recover from (\ref{mlag}) the bosonized Lagrangian 
density for the massive TW model, as  obtained by Rothe-Swieca \cite{RS}.
%%%%%%%%%%%%%%%%%%%%%%%%%%%%%%%%%%%%%%
\section{A Glance into the Non-Abelian Model}
%%%%%%%%%%%%%%%%%%%%%%%%%%%%%%%%%%%%%%

The classical Lagrangian density defining the non-Abelian TW model is given by

\be
{\cal L}\, = \,{\cal L}_{YM} \,+\,
\bar \psi\,\Big (\,i\,\gamma^\mu\,\partial_\mu\,-\,e\,
\gamma^\mu\,{\cal A}_\mu\,\Big )\,\psi\,+\,
\frac{1}{2}\,m_o^2\,\Box{tr}{\cal A}_\mu {\cal A}^\mu\,,
\ee

\no where the Yang-Mills Lagrangian is given by,

\be
{\cal L}_{YM}\,=\,-\,\frac{1}{4}\,\Box{tr}{\cal F}_{\mu \nu}{\cal F}^{\mu \nu}\,.
\ee

\no Let us apply the Wess-Zumino-Witten theory to obtain the effective bosonic
action. To this end, we shall consider the change of 
variables (\ref{cva1}), (\ref{cva2}), (\ref{fr1}), (\ref{fr2}), in 
terms of the Lie-algebra-valued Bose fields $(U,V)$. The partition function
can be factorized as \cite{BRRS,BRR},

\be
{\cal Z}\,=\,{\cal Z}_{F}^{(0)}\,{\cal Z}_{gh}^{(0)}\,\check{\cal Z}\,,
\ee

\no where ${\cal Z}_{F}^{(0)}$ is the partition function of free fermions,

\be
{\cal Z}_{F}^{(0)}\,=\,\int\,{\cal D}\,\chi\,{\cal D}\,\bar \chi\,
e^{\,i\,\int\,d ^2 z\,\bar \chi\,i\,\dslash\,\chi}\,,
\ee

\no ${\cal Z}_{gh}^{(0)}$ is the partition function of free ghosts associated 
with the change of variables (\ref{cva1})-(\ref{cva2}),

\be
{\cal Z}_{gh}^{(0)}\,=\,\int\,{\cal D}\,b_+^{(0)}\,{\cal D}\,c_+^{(0)}\,
e^{\,i\,\int\,d ^2 z\,\Box{tr} b_+^{(0)}\,i\,\partial_-\,c_+^{(0)}}\,
\int\,{\cal D}\,b_-^{(0)}\,{\cal D}\,c_-^{(0)}\,
e^{\,i\,\int\,d ^2 z\,\Box{tr} b_-^{(0)}\,i\,\partial_+\,c_-^{(0)}}\,,
\ee

\no and 

\be
\check {\cal Z}\,=\,\int\,{\cal D}\,U\,{\cal D}\,V\,e^{\,i\,S_{YM} [UV]}\,
e^{\,-\,i\,\{\,C_V\,\Gamma [UV]\,+\,\Gamma [U^{-1}]\,+\,\Gamma [V]\,\}\,-\,
\frac{1}{2 e^2}\,\{\tilde \mu_o^2 a \,+\,m_o^2\}\,
\int d ^2 z\,\Box{tr} [(U^{-1} \partial_+ U) (V \partial_- V^{-1}) ]}\,,
\ee

\no with the Yang-Mills action given by,

\be
S_{YM} [UV] = \frac{1}{4 e^2}\,\int d ^2 z\,\Box{tr} \frac{1}{2}
 [\,\partial_+ ({\G}\,i\,\partial_-\,{\G}) ]^2\,.
\ee

\no In the non-Abelian case, the WZW functional is given by \cite{WZW,AAR},

\be
\Gamma [ g ]\, =\, S_{_{P\sigma M}} [ g ]\,+\,S{_{WZ}} [ g ]\,,
\ee

\no where  $S_{_{P\sigma M}} [ g ]$ is the principal sigma model action,

\be
S_{_{P\sigma M}} [ g ]\,=\,\frac{1}{8 \pi}\,\int d ^2 x\,\Box{tr}\,
\Big [ (\partial_\mu g )\,(\partial^\mu g^{-1})\Big ]\,,
\ee

\no and  the functional $S{_{WZ}} [ g ]$ is the Wess-Zumino action,

\be
S{_{WZ}} [ g ]\,=\,\frac{1}{12 \pi}\,\int_{S_B}\,d ^3 x\,\varepsilon^{i j k}\,
\Box{tr}[(\widehat g^{-1} \partial_i \widehat g)\,(\widehat g^{-1} \partial_j \widehat g)
(\widehat g^{-1} \partial_k \widehat g)]\,.
\ee

\no Using the PW identity, the total partition function can be rewritten as,

$$
{\cal Z}\,=\,{\cal Z}_F^{(0)}\,{\cal Z}_{gh}^{(0)}\,
\int\,{\cal D}\,U\,{\cal D}\,V\,e^{\,i\,S_{YM} [UV]}\,\times
$$

\be\label{napf}
e^{\,-\,i\,\frac{1}{2 \tilde \mu_o^2}\,\{\,\tilde \mu_o^2 ( a + 2 C_V ) + m_o^2
\}\,\Gamma [UV]\,+\,i\,\frac{1}{2 \tilde \mu_o^2}\,\{ \tilde \mu_o^2 ( a - 2 )
 + m_o^2 \}\,\Gamma [V] \,
 -\,i\,\Gamma [U^{-1}]\,+\,i\,\frac{1}{2 \tilde \mu_o^2}\,\{ \tilde \mu_o^2
 \, a \, + \,m_o^2 \}\,\Gamma [U] \,\}}\,.
\ee

\no From the partition function (\ref{napf}), we read off the equivalence of
the $QCD_2$ provided with a GNI regularization $b \neq 2$ and the non-Abelian TW model
presenting a fixed bare mass $m_o^2 = \tilde \mu^2_o\,(b - a)$ for the vector field.
Similar to the Abelian case, this isomorphism also holds in 
the non-Abelian model with massive Fermi fields.

%%%%%%%%%%%%%%%%%%%%%%%%%%%%%
\section{Conclusion}
%%%%%%%%%%%%%%%%%%%%%%%%%%%%

We have re-analyzed the TW model from the functional integral approach
using the Wess-Zumino-Witten theory. The present approach give us easiness
to read off the equivalence of the $QED_2$ ($QCD_2$) with gauge symmetry breakdown and
the TW model (non-Abelian TW model). 

In the indefinite-metric formulation, the 
fields $\{\bar \psi, \psi, {\cal A}_\mu \}$ are singlet under 
gauge transformations generated by the longitudinal current. This 
implies that the field algebra is by itself the physical 
field algebra. The GI limit  can be performed
in the field operators written in terms of the non-canonical free 
field $\zeta$, leading to the Lowestein-Rothe-Swieca solution for 
the $QED_2$.

The positive-definite formulation is obtained by performing a canonical
transformation that maps the redundant Bose degrees of freedom into a
zero-norm gauge excitation. The fermion field operator is now written in
terms of the charge-carrying fermion field of the Thirring model. In this
case, the $QED_2$ limit is defined only for the GI field subalgebra. 

The ``anomalous'' \textbf{vector} Schwinger model considered in Refs. \cite{Tiao} is 
nothing but the Thirring-Wess model. As a matter of fact, the 
structural physical aspect  
underneath the conclusion given in Ref. \cite{Tiao}, according 
with the parameter $a$ apparently controls the screening and confinement 
properties in the ``anomalous'' \textbf{vector} Schwinger model, is that 
the Hilbert space of states of the TW model exhibits charge sectors 
and thus there is no confinement at all.

\no{\bf Acknowledgments}:{\it One of us (L. V. B.) wishes to thank R. L. P. G. Amaral for some
very clarifying conversations.}

%%%%%%%%%%%%%%%%%%%%%%%%%%%%%%%%%%%%%%%%%%%%%%%%%%%%%%%%%%%%%%%%

\end{document}